\documentclass[12pt]{article}

\newcommand{\m}{\mathrm}
\newcommand{\be}{\begin{equation}}
\newcommand{\ee}{\end{equation}}
\newcommand{\ba}{\begin{eqnarray}}
\newcommand{\ea}{\end{eqnarray}}

\usepackage[nosort]{cite}
\usepackage{graphicx}
\usepackage[font={scriptsize,singlespacing}]{caption}
\usepackage{epstopdf}
\usepackage{amsmath}
\usepackage{amssymb}
\usepackage{subfigure}
\usepackage{hyperref}
\usepackage{url}
\usepackage{xcolor}
\usepackage{color}

\definecolor{purple(munsell)}{rgb}{0.62, 0.0, 0.77}
\definecolor{palatinateblue}{rgb}{0.15, 0.23, 0.89}
\definecolor{royalblue(web)}{rgb}{0.25, 0.41, 0.88}
\hypersetup{linktoc=all,
    colorlinks, linkcolor={purple(munsell)},
    citecolor={palatinateblue}, urlcolor={palatinateblue}
}

\def\sideremark#1{\ifvmode\leavevmode\fi\vadjust{\vbox to0pt{\vss
 \hbox to 0pt{\hskip\hsize\hskip1em
 \vbox{\hsize2cm\tiny\raggedright\pretolerance10000
 \noindent #1\hfill}\hss}\vbox to8pt{\vfil}\vss}}}%
\begin{document}
\thispagestyle{empty}
\begin{center}

\null \vskip-1truecm \vskip2truecm

{\Large{\bf \textsf{A Note on Physical Mass and the Thermodynamics of }}}
\vskip0.2truecm
{\Large{\bf \textsf{AdS-Kerr Black Holes}}}

{\large{\bf \textsf{}}}

\vskip1truecm

{\large \textbf{\textsf{Brett McInnes}}}

\textsf{Department of Mathematics, National
  University of Singapore,\\
10, Lower Kent Ridge Road, 119076, Singapore}\\

\texttt{Email: matmcinn@nus.edu.sg}
\vskip0.3truecm
and
\vskip0.3truecm
{\large\textbf{\textsf{Yen Chin Ong}}}

\textsf{Nordita, KTH Royal Institute of Technology and Stockholm University,\\  Roslagstullsbacken 23,
SE-106 91 Stockholm, Sweden}\\
\texttt{Email:  yenchin.ong@nordita.org}

\vskip1truecm

\end{center}
\vskip1truecm \centerline{\textsf{ABSTRACT}} \baselineskip=15pt
\medskip

As with any black hole, asymptotically anti-de Sitter Kerr black holes are described by a small number of parameters, including a ``mass parameter'' $M$ that reduces to the AdS-Schwarzschild mass in the limit of vanishing angular momentum. In sharp contrast to the asymptotically flat case, the horizon area of such a black hole \emph{increases} with the angular momentum parameter $a$ if one fixes $M$; this appears to mean that the Penrose process in this case would violate the Second Law of black hole thermodynamics. We show that the correct procedure is to fix not $M$ but rather the ``physical'' mass $E=M/(1-a^2/L^2)^2$; this is motivated by the First Law. For then the horizon area decreases with $a$. We recommend that $E$ always be used as the mass in physical processes: for example, in attempts to ``over-spin'' AdS-Kerr black holes.

\newpage
\addtocounter{section}{1}
\section* {\large{\textsf{1. Black Hole Rotation Rate and Negative Cosmological Constant}}}

The four-dimensional asymptotically AdS Kerr black hole with a topologically spherical event horizon (hereinafter, the AdS-Kerr black hole) takes the form \cite{kn:carter}, in Boyer-Lindquist-like coordinates $(t,r,\theta,\phi)$,
\begin{flalign}\label{K}
g(\text{AdS-Kerr}) = &- {\Delta_r \over \rho^2}\Bigg[\,\m{d}t \; - \; {a \over \Xi}\m{sin}^2\theta \,\m{d}\phi\Bigg]^2\;+\;{\rho^2 \over \Delta_r}\m{d}r^2\;+\;{\rho^2 \over \Delta_{\theta}}\m{d}\theta^2 \\ \notag \,\,\,\,&+\;{\m{sin}^2\theta \,\Delta_{\theta} \over \rho^2}\Bigg[a\,\m{d}t \; - \;{r^2\,+\,a^2 \over \Xi}\,\m{d}\phi\Bigg]^2,
\end{flalign}
where
\begin{eqnarray}\label{eq:Q}
\rho^2& = & r^2\;+\;a^2\m{cos}^2\theta, \nonumber\\
\Delta_r & = & (r^2+a^2)\Big(1 + {r^2\over L^2}\Big) - 2Mr,\nonumber\\
\Delta_{\theta}& = & 1 - {a^2\over L^2} \, \m{cos}^2\theta, \nonumber\\
\Xi & = & 1 - {a^2\over L^2},
\end{eqnarray}
in which $L$ is the AdS length scale set by the presence of the (negative) cosmological constant; $- 1/L^2$ is the asymptotic curvature.
The parameter $a$ is the ratio of angular momentum to mass. In this work, we assume $a > 0$ without loss of generality.

The charged version of this geometry has a natural holographic interpretation in terms of a \emph{rotating} strongly coupled system on the boundary: for example, it can be used to model a rotating quark-gluon plasma. Since the QGP produced in peripheral heavy-ion collisions does indeed rotate under some circumstances \cite{1403.3258}, it is of interest to explore the properties of this spacetime more carefully. In particular, we will see that the thermodynamics of this black hole involves certain subtleties.

Note that the functions $\Xi$ and $\Delta_{\theta}$ set an upper bound for $a$ --- it should not exceed $L$. In particular, as mentioned in \cite{1504.07344}, the function $\Delta_{\theta}$ has no fixed sign \emph{a priori}, and can in principle become negative near the poles. If this happens, the signature of the spacetime in the black hole exterior would change from $(-,+,+,+)$ to $(-,+,-,-)$ as one moves from the equator to the poles, and, in particular, the signature of the conformal boundary changes from Lorentzian to Euclidean. To prevent such a (presumably unphysical) peculiarity, one \emph{has} to impose the condition that $a < L$.

The presence of the quantity $\Xi$ is necessary to ensure regularity of the (conformal) boundary metric \cite{1403.3258}. As was pointed out in that work, the spatial geometry at the boundary is approximately that of a round 2-sphere of radius $L/\sqrt{\Xi}$ near the poles. A natural question to ask is then: what happens in the limit $a \to L$, i.e., $\Xi \to 0$? It would seem that the radius of this 2-sphere will increase without bound, which is difficult to interpret\footnote{Recently, the formal limit $a \to L$ has been investigated \cite{1401.3107,1411.4309, 1504.07529} as a sort of ``solution-generating'' technique: the spacetimes so obtained are \emph{not} the AdS-Kerr spacetime any longer. Here we are interested in a different issue: how does the AdS-Kerr spacetime \emph{itself} behave when $a$ increases towards $L$?}.

\begin{figure}[!h]
\centering
\mbox{\subfigure{\includegraphics[width=3in]{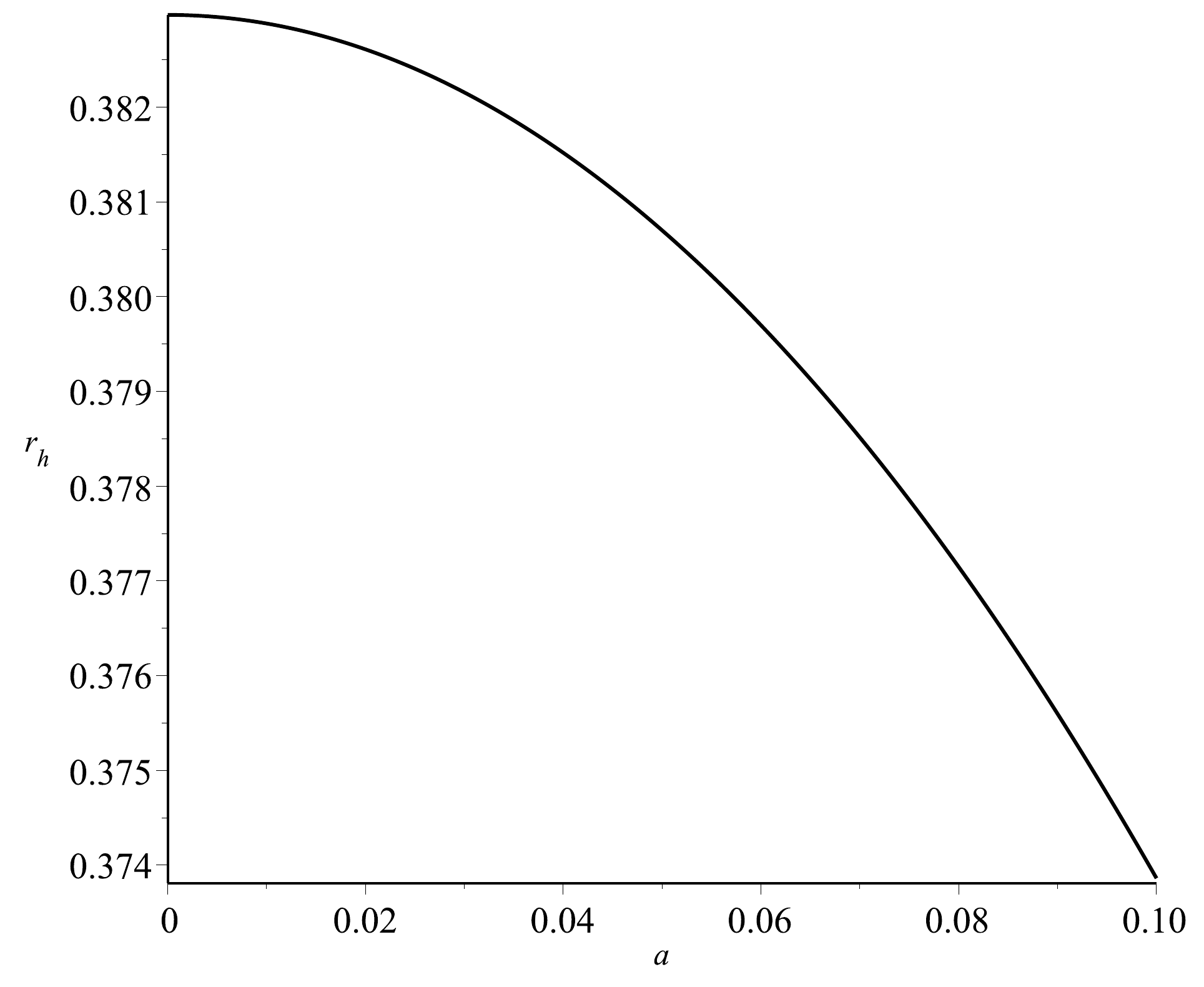}}\quad
\subfigure{\includegraphics[width=3in]{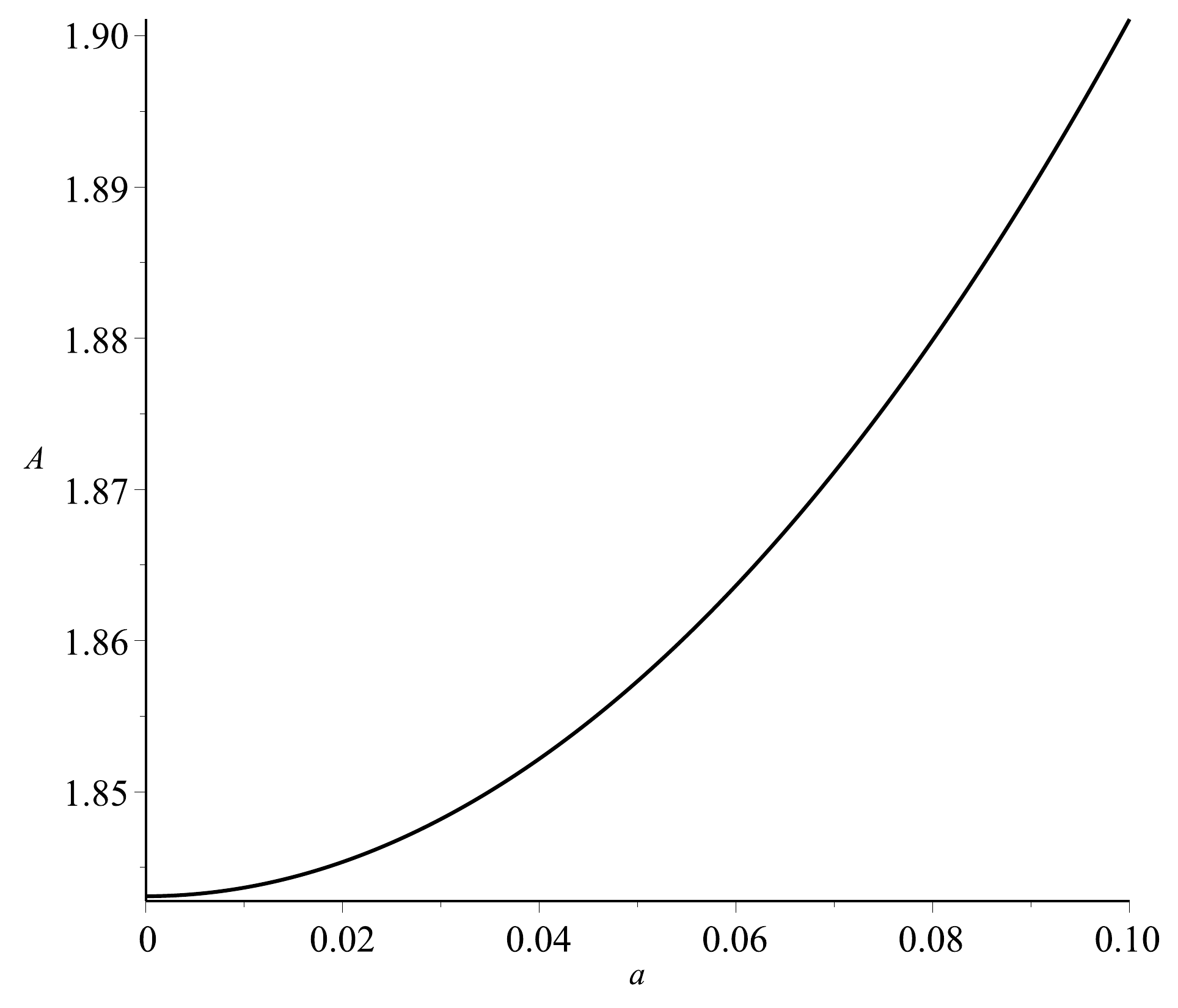} }}\caption{\label{fixedM}
\textbf{Left:} The horizon radius of Kerr-AdS black hole as a function of $a$, while keeping $M$ fixed. In this particular example, the numerical values of $L$ and $M$ are 1 and 3, respectively. \textbf{Right:} The horizon area of the same black hole as a function of $a$. The area \emph{increases} with the angular momentum parameter.}
\end{figure}

More crucially, the AdS-Kerr black hole has horizon area given by the simple expression
\begin{equation}\label{A}
A = \frac{4\pi (r_h^2 + a^2)}{\Xi},
\end{equation}
where $r_h$ denotes the value of the coordinate $r$ at the event horizon of the black hole (the largest of the two real roots of the polynomial $\Delta_r$).
This area has an apparently paradoxical property as $a$ is increased: as shown in Fig.(\ref{fixedM}), where the horizon area is graphed as a function of $a$ while keeping the mass parameter $M$ fixed, the area \emph{increases}\footnote{It is interesting to note that this does \emph{not} happen in the case of asymptotically AdS Myers-Perry black holes in odd spacetime dimensions with all angular momenta equal, as studied in \cite{grey}.} with $a$ (despite the fact that the horizon radius $r_h$ decreases with $a$).

Recall that in asymptotically flat Kerr spacetime, the area is
\begin{equation}
A = 8\pi M^2\left[1 + \sqrt{1-\left(\frac{a}{M}\right)^2}\right],
\end{equation}
which is a \emph{decreasing} function of $a$ when one fixes the mass $M$, that is, the area increases if $a$ is reduced: in fact, the area increases from 50\% to 100\% of its Schwarzschild value as $a$ decreases from values near $M$ towards zero. Energy extraction from an asymptotically flat Kerr black hole via the Penrose process \cite{penrose1, penrose2} reduces both $a$ and $M$, but the overall effect is to decrease the ratio $a/M$, and this is sufficient to overcome the tendency of the area to decrease as $M$ decreases: the area steadily \emph{increases} as the black hole spins down. As is well-known, the laws of black hole mechanics have a thermodynamical interpretation \cite{BCH, bekenstein, bekenstein2} (see \cite{0409024, 1402.5127, carlip} for some reviews) --- the horizon area corresponds to the entropy of the black hole. The increase of the area in the course of the Penrose process is thus in accord with the Second Law --- the horizon area in any classical process is always non-decreasing. Under certain hypotheses \cite{regularity} (usually including global hyperbolicity, which of course does not hold in the asymptotically AdS case) one can in fact prove mathematically that this holds.

Therefore, it is alarming that the horizon area of  a AdS-Kerr black hole \emph{increases} with $a$; for that means that, in the AdS version of the Penrose process, the area must decrease as both the mass and the angular momentum decrease; and this in turn apparently amounts to a violation of the Second Law of thermodynamics in the bulk spacetime, and, by holography, to a still more alarming violation in the dual boundary theory.

In this note, we explain in detail why this disaster does not actually happen. Essentially it is because, in order for the Second Law to hold, one must of course have a First Law. As was emphasized by Gibbons, Perry and Pope  \cite{kn:gibperry}, the validity of the First Law actually requires that we define the \emph{physical mass} of the black hole as
\begin{equation}\label{E}
E:=\frac{M}{\Xi^2},
\end{equation}
and similarly one defines a \emph{physical angular momentum},
\begin{equation}\label{J}
J:=\frac{aM}{\Xi^2}=aE.
\end{equation}
Notice that the parameter $a$ is the ratio of the physical angular momentum to the physical mass. Therefore, the analysis above is unphysical since we were fixing the wrong quantity, $M$.
Note also that the physical (Abbott-Deser \cite{AD}) mass depends on $a$. See also \cite{HT}. It is, of course, also the correct \emph{conserved quantity} corresponding to the timelike Killing vector of the geometry \cite{0506057}.

Despite the fact that the physical quantities --- as identified by the First Law and the Killing symmetry --- are already known, there is still some confusion in the literature. For example, it has been claimed that AdS-Kerr black holes could be over-spun, that is, be transformed into naked singularities. If true, this would be a violation of the cosmic censorship conjecture \cite{gao}. In considering a physical process such as spinning-up a black hole, however, it is crucial that one uses the correct, physical, quantities; and not a quantity like $M$ that would lead to a violation of the Second Law (and therefore would mean that said process does not qualify as being \emph{physical}.)

In this work, we will show explicitly that the Second Law is maintained \emph{if} we hold $E$ fixed instead of $M$. (Note that once we hold $E$ fixed, varying the angular momentum $J$ is the same as varying $a$.)  Furthermore, we propose that the cosmic censorship conjecture continues to hold if one works with the physical mass $E$ instead of $M$. While it is conceivable that the cosmic censorship can be violated under some circumstances, naked singularities should not be generic occurrences in general relativity \cite{hod}, especially not in the context of AdS-Kerr black holes, as we will further discuss in the conclusion.

\addtocounter{section}{2}
\section* {\large{\textsf{2. Some Properties of Kerr-AdS Black Holes}}}

Cosmic censorship for these black holes is usually expressed \cite{9808097} using $M$: one needs to have
\begin{equation}
\frac{M}{L} \geqslant {\Gamma}(a/L):=\frac{1}{3\sqrt{6}}\left[\sqrt{1+\frac{14 a^2}{L^2} +\frac{a^4}{L^4} }+ \frac{2a^2}{L^2} +2\right] \times \left[\sqrt{1+\frac{14 a^2}{L^2} +\frac{a^4}{L^4} } - \frac{a^2}{L^2} -1\right]^{\frac{1}{2}}.
\end{equation}
One can of course formulate this condition using $E$ instead of $M$:
\begin{equation}
\frac{E}{L} \geqslant {\beth}(a/L):=\frac{1}{3\sqrt{6}~{\Xi^2}}\left[\sqrt{1+\frac{14 a^2}{L^2} +\frac{a^4}{L^4} }+ \frac{2a^2}{L^2} +2\right] \times \left[\sqrt{1+\frac{14 a^2}{L^2} +\frac{a^4}{L^4} } - \frac{a^2}{L^2} -1\right]^{\frac{1}{2}}.
\end{equation}
One may check that in the limit $L \to \infty$ both of these recover the well-known censorship condition for asymptotically flat Kerr black holes, namely, $M \geqslant a$.

\begin{figure}[!h]
\centering
\mbox{\subfigure{\includegraphics[width=3in]{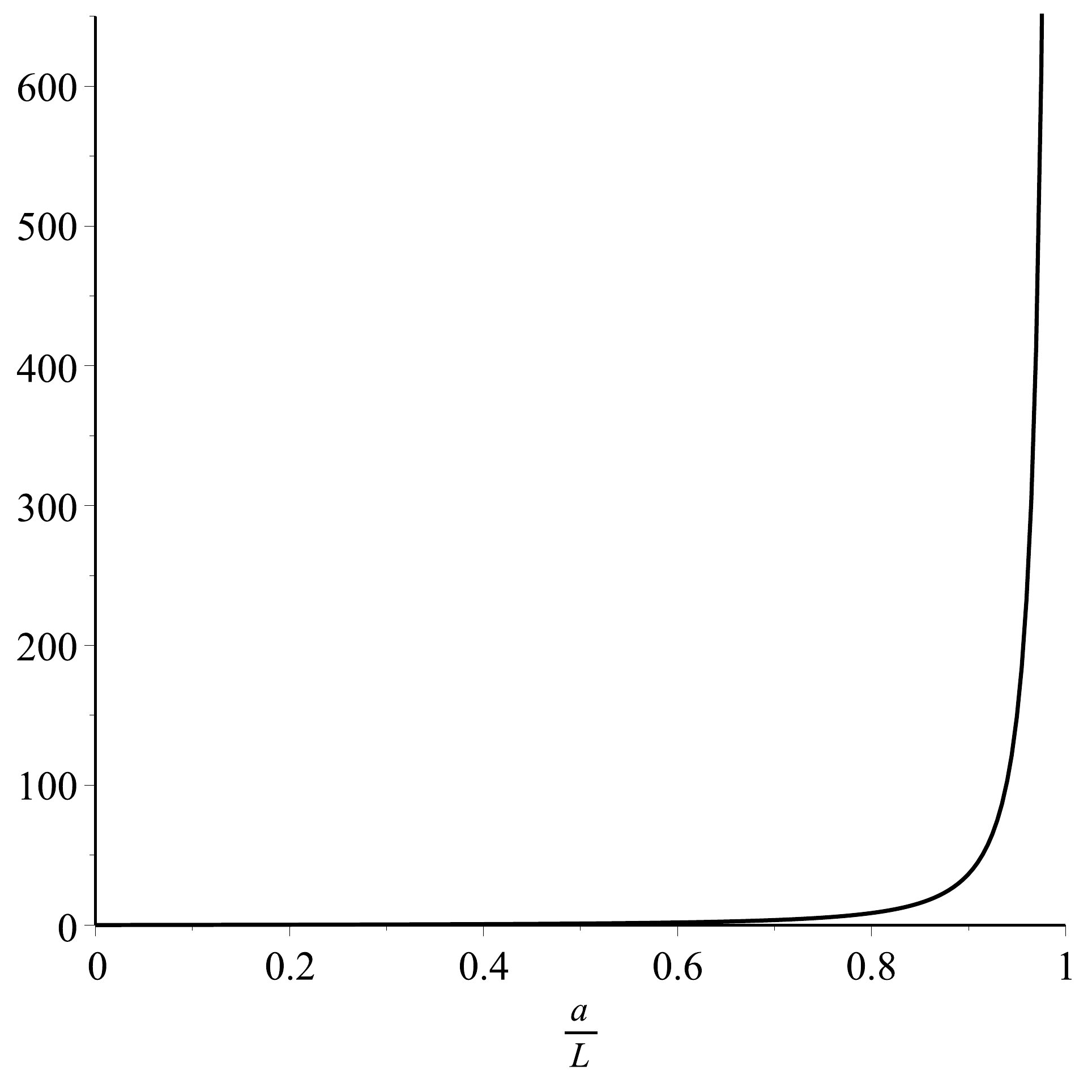}}\quad
\subfigure{\includegraphics[width=3in]{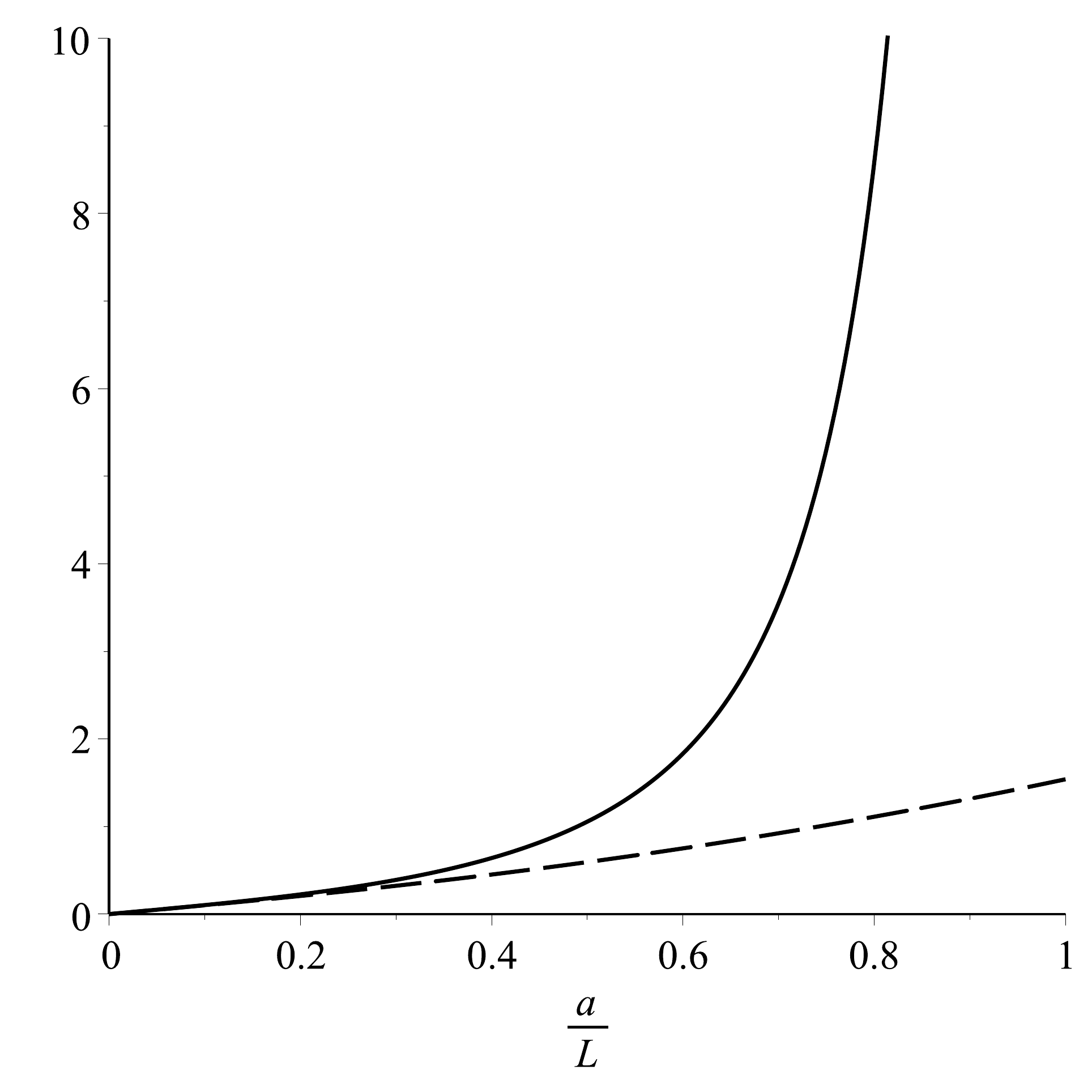} }}\caption{\label{beths}
\textbf{Left:} The function $\beth$ asymptotes to the vertical line $a/L=1$. Black hole solutions are only allowed above this graph. \textbf{Right:} The function $\beth$ (solid) compared to $\Gamma = \beth \Xi^2$(dashed).}
\end{figure}

Plots of $\beth (a/L)$ and $\Gamma (a/L)$ against $a/L$ are shown in Fig.(\ref{beths}). We see that the two functions differ significantly as $a \to L$.
The function $\beth (a/L)$ is the one that is important here --- AdS-Kerr black holes only exist for pairs ($a/L$, $E/L$) above the graph of $\beth (a/L)$. Note that, in particular, $\beth (a/L)$ asymptotes to the vertical line $a/L=1$, so we see at once that, \emph{in any process such that the physical mass remains finite, censorship precludes the possibility that $a$ will ever reach $L$}. In particular, if we fix $E$, then, moving horizontally in the figure, we eventually hit the graph at some value of $a/L$ which is strictly less than unity. Again, in the asymptotically flat case, cosmic censorship only requires that $a/M < 1$, allowing values arbitrarily close to unity; here, $a/M < 1$ still holds, but
\begin{equation}
a^* := \frac{a}{E} = \frac{a}{M}\left(1-\frac{a^2}{L^2}\right)^2< \frac{a}{M} < 1.
\end{equation}
That is, cosmic censorship is ``stricter'' for AdS-Kerr black holes, in the sense that $a^*$ has to be smaller than some value strictly less than unity. (See \cite{1108.6234} for a discussion in terms of $M$ and $\Gamma (a/L)$.)

\begin{figure}[!h]
\centering
\includegraphics[width=3in]{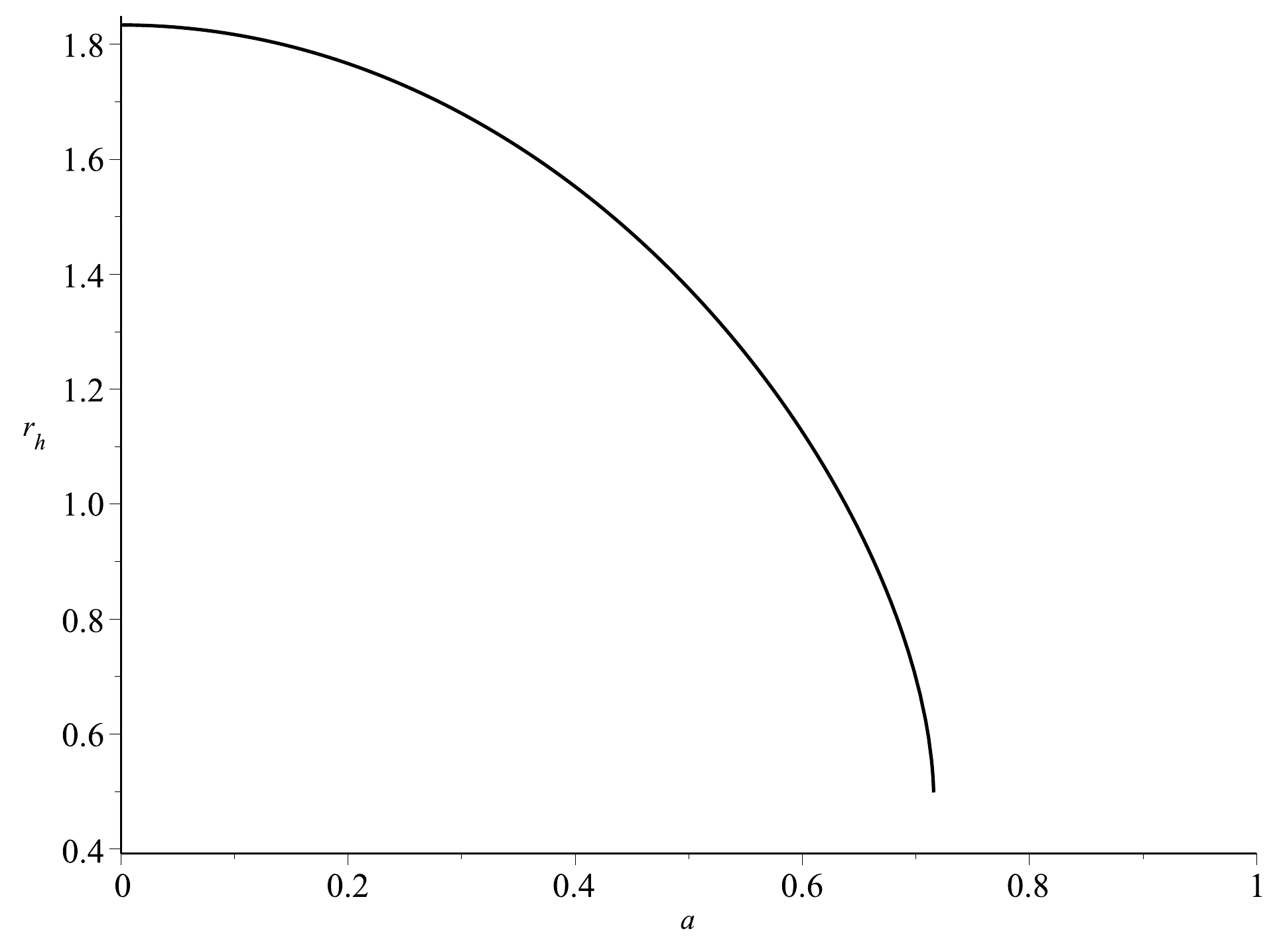}
\caption{\label{ra} The horizon radius $r_h$ as a function of $a$, with fixed $E$. In this example, $L=1$ and $E=4$.}
\end{figure}

In asymptotically flat Kerr geometry, $a^*$ measures how close the black hole is to extremality. Specifically, $a^*=0$ means the black hole is not rotating (a Schwarzschild black hole), while $a^*=1$ means that we have an extremal black hole. However, as emphasized in \cite{1108.6234}, asymptotically flat spacetime intuition can be misleading when working with asymptotically AdS spacetimes, and $a^*$ here can in fact be small even if the black hole is near-extremal. In fact, an example was provided in which a Kerr-AdS black hole with $a^* \approx 0.177$ was shown to be rotating at 98.8\% of its extremal value.
This black hole, for fixed numerical value $L=1$, corresponds to fixing $E\approx 4$ in the same unit of length. One could plot the horizon radius $r_h$ as a function of $a$ and note that indeed the graph terminates at around $a\approx 0.72 L < L$ (consistent with the value calculated in \cite{1108.6234}). This is shown in Fig.(\ref{ra}).

\begin{figure}[!h]
\centering
\includegraphics[width=3in]{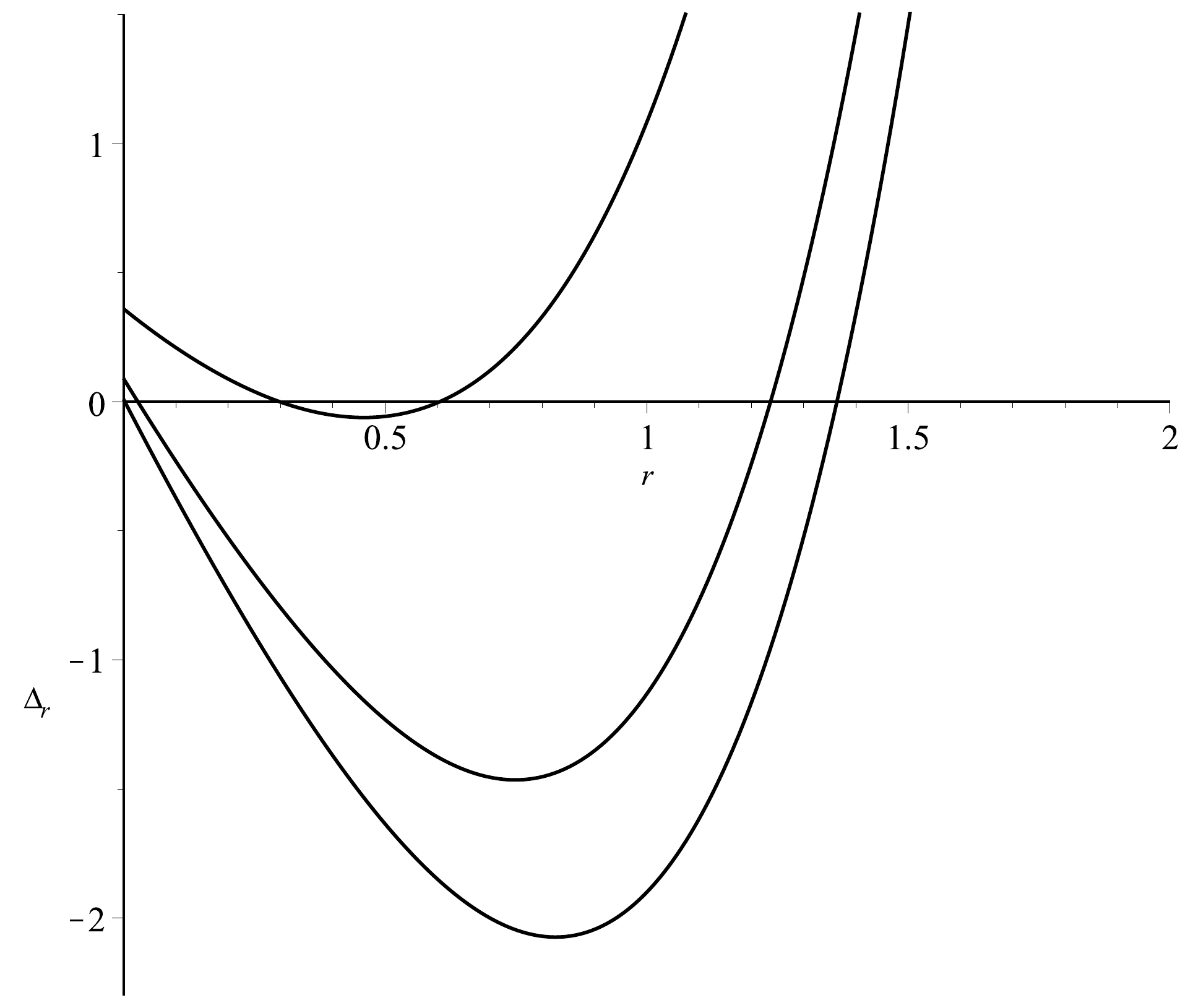}
\caption{\label{Deltar} The polynomial $\Delta_r$ with various values of the rotation parameter $a$. For any fixed $a$, the larger root corresponds to the event horizon. In this example, we use the numerical values $L=1$, $E=2$, and from right to left, $a=0.1, 0.3, 0.6$, respectively.}
\end{figure}

We notice from Fig.(\ref{ra}) that $r_h$ is a decreasing function of $a$. This is also clear if one examines the polynomial $\Delta_r$. In Fig.(\ref{Deltar}), we plot $\Delta_r$ and see that it has two roots --- an event horizon and an inner Cauchy horizon. With increasing $a$, the value of the bigger root (the event horizon) decreases, while the value of the smaller root (the Cauchy horizon) increases, until they meet at the extremal value. See also \cite{0804.3811}.

It is in fact not difficult to prove that the (event) horizon radius $r_h$ must be a decreasing function of $a$. Upon differentiating the polynomial $\Delta_r$ with respect to $a$ and rearranging terms, we can obtain
\begin{equation}
\frac{\partial r_h}{\partial a} = -\left[\frac{2a + \frac{2a r_h^2}{L^2} + \frac{8aE}{L^2}\left(1-\frac{a^2}{L^2}\right)r_h}{2r_h + \frac{4r_h^3}{L^2} + \frac{2r_ha^2}{L^2}-2E\left(1-\frac{a^2}{L^2}\right)^2}\right].
\end{equation}
The numerator of this expression is positive; while the denominator is none other than $\partial \Delta_r/\partial r$ evaluated on the horizon. This derivative is non-negative --- it approaches zero in the extremal limit when the two real roots tend together. Therefore  $\partial r_h/\partial a$ is negative, and so the horizon radius decreases with $a$. Note that this result does not depend on whether one fixes $E$ or fixes $M$. This will no longer be the case with the area.

\addtocounter{section}{3}
\section* {\large{\textsf{3. The Second Law for AdS-Kerr Black Holes}}}

The area of the event horizon of a Kerr-AdS black hole is
\begin{equation}
A = \dfrac{4\pi (r_h^2 +a^2)}{\Xi} = \dfrac{4\pi}{\Xi}\dfrac{2Mr_h}{1+\frac{r_h^2}{L^2}} = \frac{8\pi E \left(1-\frac{a^2}{L^2}\right)r_h}{1+\frac{r_h^2}{L^2}}.
\end{equation}

With fixed $E$, we first note that $1-a^2/L^2$ is a decreasing function of $a$, whereas the function
\begin{equation}
\frac{r_h}{1+\frac{r_h^2}{L^2}}=\left(\frac{1}{r_h} + \frac{r_h}{L^2}\right)^{-1}
\end{equation}
is a decreasing function of $a$ if and only if
\begin{equation}
\mathcal{F}(L,a,r_h):=\frac{1}{r_h} + \frac{r_h}{L^2}
\end{equation}
is an increasing function of $a$.
A straightforward calculation yields
\begin{equation}
\frac{\partial \mathcal{F}}{\partial a} = \frac{\partial r_h}{\partial a} \left[\frac{r_h^2 - L^2}{r_h^2 L^2}\right].
\end{equation}
Since $\partial r_h /\partial a < 0$, we see that $\mathcal{F}$ is an increasing function of $a$ if and only if $r_h \leqslant L$.  That is, the horizon areas of ``small'' Kerr-AdS black holes decrease with $a$ if we fix $E$.

\begin{figure}[!h]
\centering
\includegraphics[width=3in]{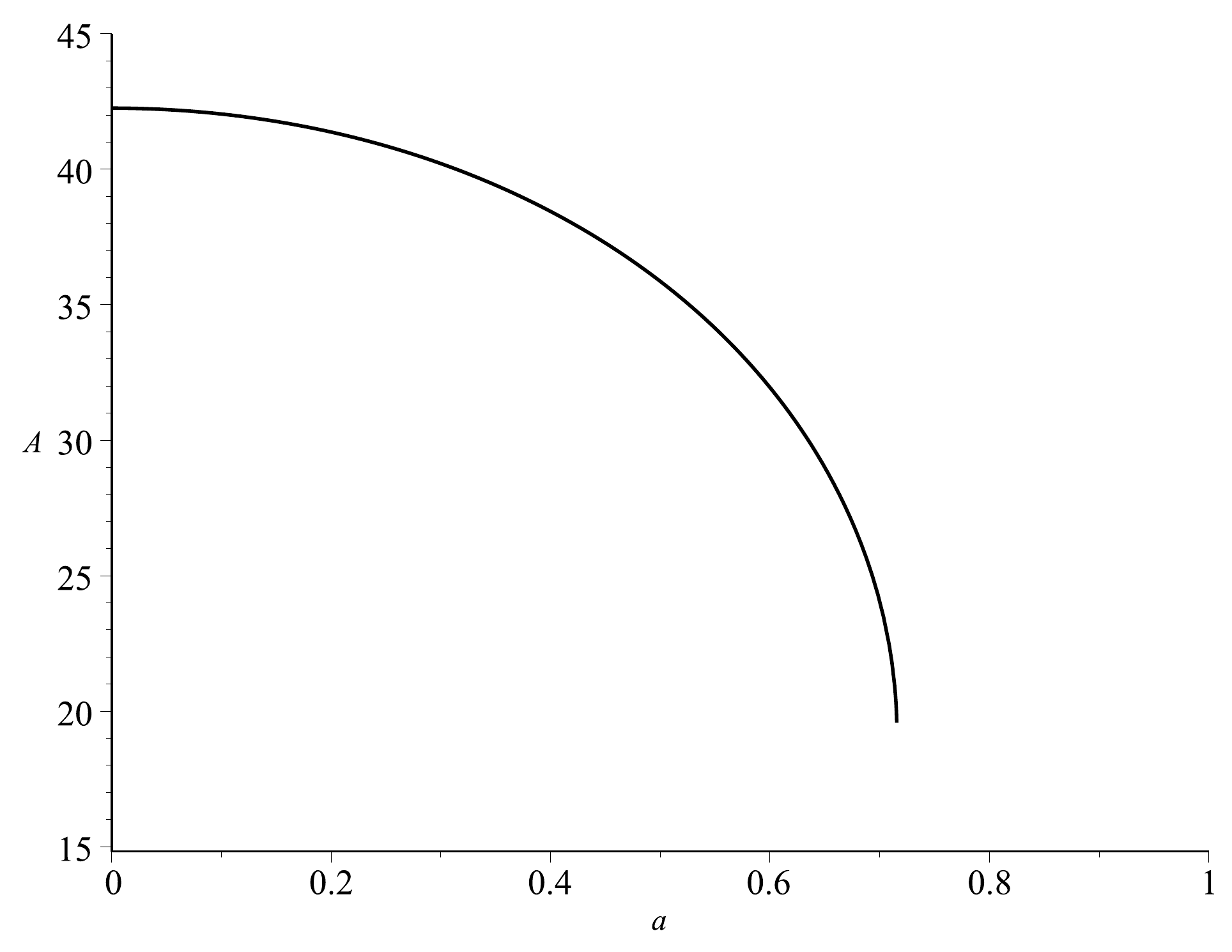}
\caption{\label{area} The horizon area $A$ is a decreasing function of $a$ if we fix $E$. In this example, $L=1$ and $E=4$.  The horizon radius of the same black hole was plotted in Fig.(\ref{ra}).}
\end{figure}

The same result is actually also true for ``large'' AdS-Kerr black holes (that is, black holes with $r_h \geqslant L$). To prove this analytically, one should examine the derivative $\partial A /\partial a$ explicitly.
It is given by
\begin{equation}
\frac{1}{8\pi E}\frac{\partial A}{\partial a} =\left(1+\frac{r_h^2}{L^2}\right)^{-1} \left(L^2+r_h^2\right)^{-1} r_h^2\left[\left(1-\frac{a^2}{L^2}\right)\left(\frac{L^2}{r_h^2}-1\right)\frac{\partial r_h}{\partial a} - \frac{2a}{r_h}\right],
\end{equation}
so the sign of this derivative depends on the sign of the function
\begin{equation}
\mathcal{G}(L,a,r_h):=\left(1-\frac{a^2}{L^2}\right)\left(\frac{L^2}{r_h^2}-1\right)\frac{\partial r_h}{\partial a} - \frac{2a}{r_h}.
\end{equation}
However, as with many calculations involving rotating black holes, analytic proofs are at best tedious and not particularly illuminating. (Note that for ``small'' AdS-Kerr black holes, $\mathcal{G}$ is indeed negative.) We therefore approach the issue numerically. An explicit example is provided by Fig.(\ref{area}), which corresponds to the same black hole whose horizon radius was plotted in Fig.(\ref{ra}).

\begin{figure}[!h]
\centering
\mbox{\subfigure{\includegraphics[width=3in]{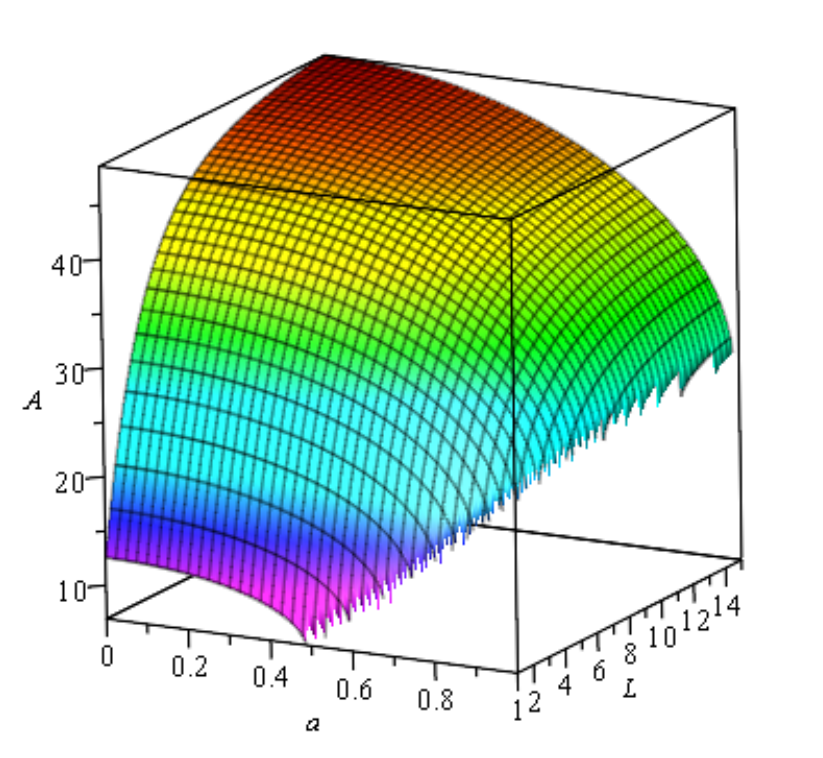}}\quad
\subfigure{\includegraphics[width=3in]{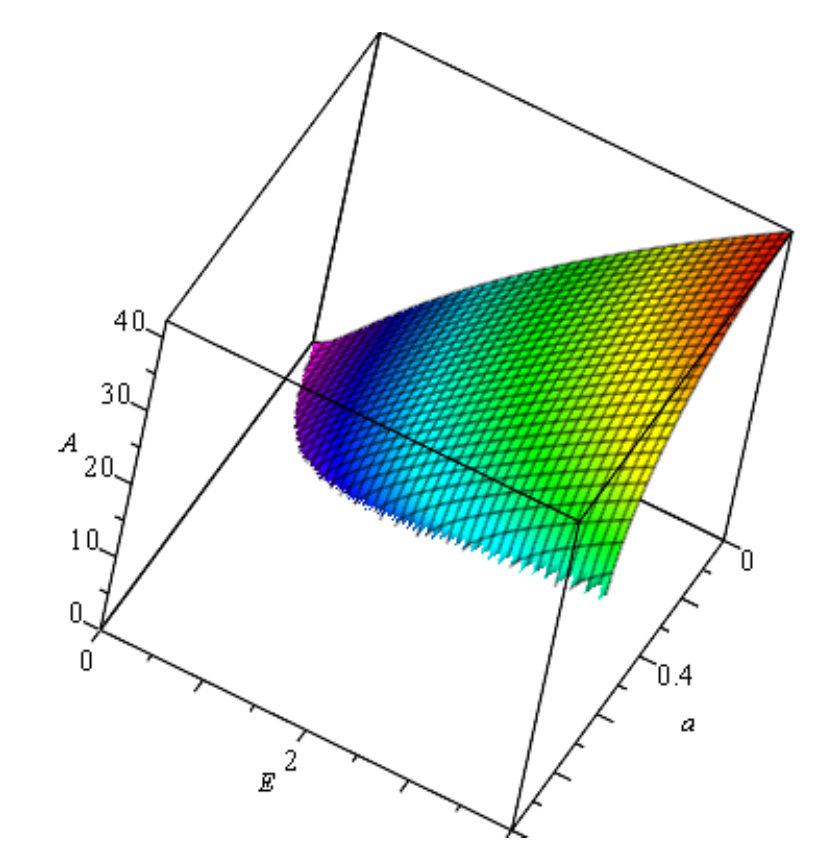} }}
\caption{\label{3D} \textbf{Left:} A plot of the horizon area of Kerr-AdS black hole with fixed physical mass, here with $E=1$. \textbf{Right:} A plot of the horizon area of Kerr-AdS black hole with fixed $L$, here with $L=1$.}
\end{figure}

One can see how the area changes if $E$ is fixed and $L$ is varied and vice versa from Fig.(\ref{3D}). For fixed physical mass $E$ (left panel), The effect of increasing the value of the AdS length scale $L$ is to increase the allowed values of the rotation parameter $a$ before the black hole loses its event horizon (and thus ceases to be a black hole), but the qualitative features otherwise remain the same. Fixing $L$ and varying the value of $E$ produces the graph in the right panel. In a Penrose process, both $a$ and $E$ would decrease, but the shape of the surface is such that it is possible for the corresponding trajectory to rise up, as they do so, to larger values of $A$. \emph{In short, such a process respects the Second Law} in the asymptotically AdS case, just as in the asymptotically flat case.

\newpage
\addtocounter{section}{4}
\section* {\large{\textsf{4. Conclusion: Physical Mass and the Fate of AdS}}}

\emph{A priori}, at the level of pure mathematics, there is nothing wrong with fixing the mass parameter $M$ instead of the physical mass $E$, and study the properties of the spacetime geometry under such an assumption. However, we have seen that using the appropriate definition of physical mass is crucial for preserving the Second Law as it applies to AdS-Kerr black holes. It is also crucial for that purpose in another sense, as we now discuss.

The non-linear instability of anti-de Sitter spacetime \cite{bizon} is a phenomenon of potentially great importance for applications of holography \cite{horror} (see \cite{dias} for a recent interesting discussion). The suggestion is that even a small perturbation of AdS will lead to the formation of a black hole, which will generically be of the AdS-Kerr type. However, cosmic censorship has been questioned
\cite{gao} in precisely this context. If censorship really breaks down for AdS-Kerr black holes, then the non-linear instability would mean that \emph{a non-negligible class of  perturbations of AdS can lead to naked singularities}, leading to a breakdown of the concept of black hole entropy, and to very unwelcome consequences for the dual theory on the boundary. Therefore, although it is ``formally'' not incorrect to fix the mass parameter $M$, one sees that it is not the correct quantity that one should use to deal with \emph{physical} processes, such as those that involve either energy extraction or over-spinning of the black holes.

The usual way of probing censorship in this context is to try to ``overspin'' the black hole\footnote{For such attempts in various asymptotically flat black hole geometries, see \cite{mariam}.}: one imagines projecting a particle of mass $\varepsilon$ and angular momentum $\lambda$ into an AdS-Kerr black hole with parameters ($M, a$). One is tempted to claim that the resulting black hole has parameters ($M',a'$) = ($M + \varepsilon, {aM+\lambda\over M+\varepsilon}$). However, this would lead to a violation of the cosmic censorship conjecture \cite{gao}, which is a great concern as we explained above.
However, according to our discussion here, \emph{this procedure is not correct}. Instead one has a more complicated algebraic problem to solve in order to determine $M'$ and $a'$, as follows. Defining $\Xi'$ in the obvious way, one has, from our discussion in Section 1 above,
\begin{equation}\label{MPRIME}
{M'\over \Xi^{'2}} = {M\over \Xi^2} + \varepsilon
\end{equation}
and, using this, we have
\begin{equation}
a' = {{aM\over \Xi^2}+\lambda\over {M\over \Xi^2} + \varepsilon}.
\end{equation}
Regarding $M, a, \varepsilon,$ and $\lambda$ as known, we can use this relation to compute $a'$; then $\Xi'$ is known, and equation (\ref{MPRIME}) can be used to compute $M'$; clearly the relation between ($M, a$) and ($M',a'$) is far from simple. We conjecture that, if one proceeds in this manner, one will now find that it is \emph{not} possible to overspin these black holes\footnote{The attempt to overspin Myers-Perry-AdS black holes with equal angular momenta in odd dimensions has been made \cite{1402.4840}; it was shown that cosmic censorship does indeed hold in these cases.}$^{,}$\footnote{After our work appeared on the arXiv, Gwak and Lee \cite{1509.06691} have since shown that AdS-Kerr black holes indeed cannot be over-spun if one works with the physical mass $E$.}. This is motivated, as mentioned above, by holography: such objects should always have a well-defined entropy, that is, they should always have an event horizon, because the entropy of the boundary theory is always well-defined. The basic observation here is, once more, that AdS-Kerr thermodynamics only works if one adopts the correct definition of the physical mass.

\begin{figure}[!h]
\centering
\includegraphics[width=3in]{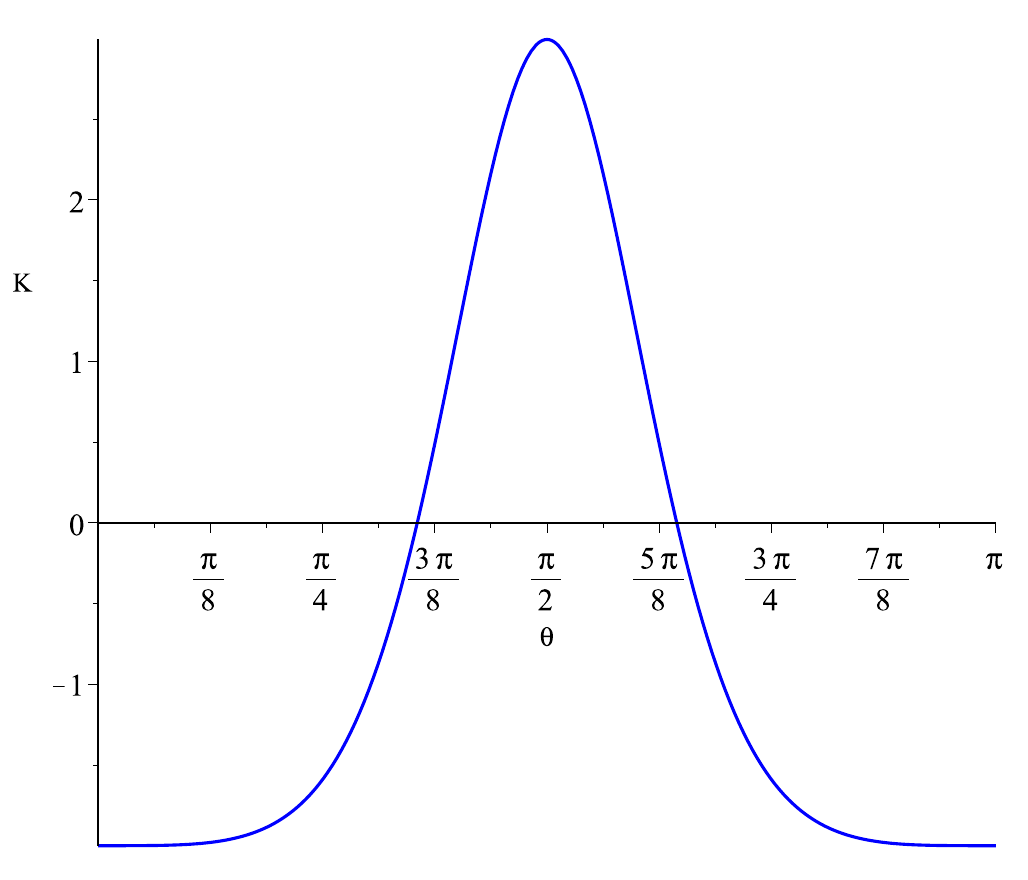}
\caption{The Gaussian curvature, K, of the event horizon of a Kerr-AdS black hole, plotted against the polar angle $\theta$. In this example, $a^2=0.999$, $M=2$ and $L=1$. \label{7}}
\end{figure}

However, all this is \emph{not} to say that the parameter $M$ is of no interest. On the contrary, it plays a definite role in determining the geometry of the black hole spacetime. To take but one example: consider an AdS-Kerr black hole with a very large physical mass,  $E = 2 \times 10^6$, in units such that $L = 1$, and with an almost equally large angular momentum such that $a^2 = 0.999$. The geometry here is very different to that of a similarly massive asymptotically flat Kerr black hole: for example, the (outer) event horizon is located at quite a small value of the radial coordinate, $r_h \approx 1.0004998$. In view of this, one might expect that the geometry of the event horizon would be severely distorted (see \cite{kn:visser} for a discussion of the geometry of the event horizon in the asymptotically flat case). However, that is not so: the Gaussian curvature of the event horizon is quite small: see Fig.(\ref{7}).

(The fact that the Gaussian curvature is negative near the poles is not surprising: see \cite{kn:visser}.) Clearly, the fact that the physical mass is enormous, and that the event horizon is not ``far away'', gives a very misleading impression of the geometry here. On the other hand, the mass parameter $M$ is just equal to 2 in these units in this spacetime. Evidently $M$ gives a much better indication of the geometry than the physical mass.

\addtocounter{section}{5}
\section*{\large{\textsf{Acknowledgement}}}
BMc is grateful to Dr. Soon Wanmei, J.L. McInnes, and C.Y. McInnes, for helpful discussions.
YCO thanks the Yukawa Institute for Theoretical Physics for hospitality, where part of this work was completed. He also thanks Stanley Deser for discussions.

\end{document}